\documentclass[a4paper,10pt,twoside]{cpc-hepnp}
\usepackage{CJK,upgreek,fancyhdr}
\usepackage{multicol}
\usepackage{graphicx}
\usepackage{booktabs}
\usepackage{xcolor, soul}
\sethlcolor{yellow}
\usepackage{amssymb,bm,mathrsfs,bbm,amscd}
\usepackage[tbtags]{amsmath}
\usepackage{lastpage}
\usepackage{subcaption}

\begin{document}
\begin{CJK*}{GB}{gbsn}

\fancyhead[c]{\small Chinese Physics C~~~Vol. xx, No. x (201x) xxxxxx}
\fancyfoot[C]{\small 010201-\thepage}

\footnotetext[0]{Received 31 June 2015}

\title{Spectroscopic Study of Strangeness=-3 $\Omega^{-}$ Baryon}

\author{%
      Chandni Menapara, \email{chandni.menapara@gmail.com}%
and
 Ajay Kumar Rai
}
\maketitle

\address{%
Department of Physics, Sardar Vallabhbhai National Institute of Technology, Surat-395007, Gujarat, India\\
}

\begin{abstract}
$\Omega^{-}$ baryon with sss quarks has been scarcely observed in the experiments so far and has been studied through many theoretical studies only. Here, an attempt has been made to explore properties of $\Omega$ with hypercentral Constituent Quark Model (hCQM) with a linear confining term. The resonance mass spectra has been obtained for 1S-4S, 1P-4P, 1D-3D and 1F-2F respectively. The Regge trajectory has been investigated for the linear nature based on calculated data alongwith the magnetic moment. The present work has been compared with various approaches and known experimental findings. 
\end{abstract}


\begin{pacs}
1 -- 3 PACS codes (Physics and Astronomy Classification Scheme, https://publishing.aip.org/publishing/pacs/pacs-2010-regular-edition/)
\end{pacs}

\footnotetext[0]{\hspace*{-3mm}\raisebox{0.3ex}{$\scriptstyle\copyright$}2013
Chinese Physical Society and the Institute of High Energy Physics of the Chinese Academy of Sciences and the Institute of Modern Physics of the Chinese Academy of Sciences and IOP Publishing Ltd}%

\begin{multicols}{2}

\section{Introduction}
\label{intro}
The discovery of $\Omega$ baryon dates back to 1964 and yet till today it is the least observed in the experiments worldwide. $\Omega^{-}$ holds a place in decuplet family with isospin I=0 and strangeness S=-3 with sss quarks. 
\begin{table*}
\centering
\caption{PDG $\Omega$ baryon \cite{pdg}}
\label{tab:1 pdg}
\begin{tabular}{ccc}
\hline
State & $J^{P}$ & Status \\
\hline
$\Omega$(1672) & $\frac{3}{2}^{+}$ & **** \\
$\Omega$(2012) & $?^{-}$ & *** \\
$\Omega$(2250) &  & *** \\
$\Omega$(2380) & & ** \\
$\Omega$(2470) & & ** \\
\hline
\end{tabular}
\end{table*}
The search for missing resonances is the prime aim of the hadron spectroscopy so as to understand the internal dynamics of the quarks inside the system ranging from light hadrons to heavy as well as exotic hadrons as depicted by few recent reviews\cite{theil, chen}. The motivation behind the current study is to exploit all resonance mass with possible spin-parity assignment. This is the extension of the previous study for non-strange \cite{z19,c1} and strange baryons with S=-1,-2 \cite{c2}.\\

$\Omega$ baryon belongs purely to the decuplet representation in the similar manner as $\Delta$, while $\Sigma$ and $\Xi$ can in principle be realized as a mixture of octet and decuplet states. In terms of multi-strangeness, $\Omega$ baryon is in a similar position like $\Xi$ as both are not easily observed in experiments and not more information has been readily available since bubble chamber data. n a recent study of strong decays in costituent quark model with relativisitc corrections, it is highlighted that unlike other light baryons, $\Omega$ with only strange quarks may be a tool to reach the valence quarks in the baryons when other kaon cloud effects are somehow excluded \cite{arifi}. Pervin {\it et al.} \cite{pervin} has vividly described that  multi-strange baryons are produced only as a part of final state and also with very small production cross-section, making the analysis more complicated to study them. Recent studies at Belle experiments have provided with some results as $\Omega$(2012) through $e^{+}e^{-}$ annihilations and into $\Xi^{0}K^{-}$ as well as $\Xi^{-}\bar K^{0}$ decay channels \cite{belle}. Earlier, BaBar collaboration attempted to study the spin of $\Omega^{-}$(1672) for $J=\frac{3}{2}$ through processes like $\Omega_{c}^{0}\rightarrow \Omega^{-} K^{+}$  and $\Xi_{c}^{0}\rightarrow \Omega^{-} \pi^{+}$\cite{babar}.  All these observations pose a challenge towards the underlying mystery of multi strange baryons especially for S=-3. The upcoming experimental facility at FAIR, $\bar P$ANDA-GSI is expected to perform a dedicated study of hyperons especially at low energy regime \cite{panda} as well as a part of BESIII experiment shall be including the strange quark systems \cite{BESIII} and J-PARC facility \cite{jparc}. \\

The three star state $\Omega$(2012) has been a puzzling one appearing in table \ref{tab:1 pdg} as the second known state. The discovery of $\Omega$(2012) by Belle collaboration sparked a lot of theoretical work on the issue, with pictures inspired by quark models as well as molecular pictures based on the meson-baryon interaction. In various quark models, the masses of the first orbital excitations of  states were deemed to $\Omega$(2012). A recent study has proposed this state to be a molecular one. This state is slightly below $\Xi$(1530){$\bar K$} threshold so that the binding mechanism could be a coupled channel dynamics \cite{polyakov}. Its characteristic signature could be a three body channel $\Xi${$\bar K$}$\pi$. There are other studies which signify that present information is not enough to consider it as a molecular state \cite{pavon,lin} as well as some other disapproving the proposed state \cite{Belle19}.  Also, a study has been revisited to check the compatibility of molecular picture of 2012 within the coupled channel unitary approach \cite{ikeno}.  Xiao {\it et al.} has studied the strong decays within chiral quark model to understand the structure of $\Omega$(2012) state \cite{xiao}. There are several models to study the $\Omega$ baryon properties through theoretical and phenomenological such as quark pair-creation \cite{wang}, QCD Sum rule \cite{aliev}, Glozman-Riska model \cite{riska}, algebraic model by Bijker \cite{bijker} and large-$N_{c}$ analysis \cite{matagne,goity}. Recently A. Arifi {\it et al.} has investigated the decay properties of $\Xi$ and $\Omega$ baryons including Roper-like resonances using relativistic corrections in constituent quark model \cite{arifi}. Our group has also attempted to explore light, strange baryon through Regge Phenomenology \cite{juhi}. Nambu-Jona-Lasinio (NJL) approach has been used for calculating the mixing of three and five components in low-lying $\Omega$ states with negative parity \cite{an}. The partial wave analysis of light baryons is also very important tool for the spectroscopy of narrow experimental states \cite{qin, miguel, liang, azimov}. Also, the details of these models through comparison are described in section 3. \\

In the present article, we study the resonance mass spectra of $\Omega^{-}$ baryon through a non-relativistic model. The section II describes the potential terms utilized to get the resonance mass with spin-dependent and correction parts. The section III sketches the results obtained through the model as well as summarizes the comparison with other models. Sections IV and V exploits the Regge trajectories and magnetic moment properties leading to the conclusion of the study.

\section{Theoretical Background}
\label{sec:1}

\begin{figure*}

\centering
\caption{\label{fig:Jacobi} Representation of three-body system \cite{sattari}}
\includegraphics[scale=1.0]{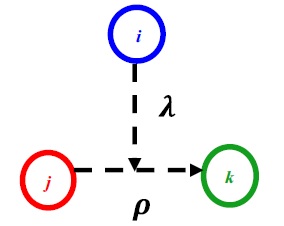}
\end{figure*}

The present study is based on the hypercentral Constituent Quark Model (hCQM), a non-relativistic approach \cite{rai,zalak}. The baryons are composed of three quarks confined within and interacting by a potential which here is considered to be hypercentral. The hyperspherical coordinates are given by the angles $\Omega_{\rho}$,  $\Omega_{\lambda}$ along with hyperradius x and hyperangle $\xi$ which are written in terms of Jacobi coordinates as in figure \ref{fig:Jacobi} \cite{ferraris,giannini},
\begin{subequations}
\begin{align}
 {\bm{\rho}} = \frac{1}{\sqrt{2}}({\bf r_{1}} -{\bf r_{2}}) \\
{\bm{\lambda}} = \frac{(m_{1}{\bf r_{1}} + m_{2}{\bf r_{2}} - (m_{1}+m_{2}){\bf r_{3}})}{\sqrt{m_{1}^{2} + m_{2}^{2} + (m_{1}+m_{2})^{2}}} 
\end{align}
\end{subequations}
The $r_{i}$ and $m_{i}$ is the internal distance between given two quarks and their masses respectively.
The hyperradius and hyperangle are defined as 
\begin{equation}
x = \sqrt{{\bf \rho^{2}} + {\bf \lambda^{2}}}
; \; \; \xi = arctan(\frac{\rho}{\lambda})
\end{equation} 
 The hyperradius x is a one-dimensional coordinate that encloses the effects of the three-body interaction at the same time. The quarks are pictured to be connected by gluonic strings and the potential increases linearly with the radius x.
The reduced masses with Jacobi co-ordinates $\bm{\rho}$ and $\bm{\lambda}$ given by
\begin{equation}
m_{\rho}=\frac{2m_{1}m_{2}}{m_{1}+m_{2}}; \; \; m_{\lambda}=\frac{2m_{3}(m_{1}^{2}+ m_{2}^{2}+m_{1}m_{2})}{(m_{1}+m_{2})(m_{1}+m_{2}+m_{3})}
\end{equation}
The kinetic energy operator in the center of mass frame is written as
\begin{equation}
-\frac{\hbar^{2}}{2m}(\Delta_{\rho} + \Delta_{\lambda})= \frac{\hbar^{2}}{2m}(\frac{\partial^{2}}{\partial x^{2}} + \frac{5}{x}\frac{\partial}{\partial x} + \frac{L^{2}(\Omega)}{x^{2}})
\end{equation}
Here, $L^{2}(\Omega)= L^{2}(\Omega_{\rho}, \Omega_{\lambda}, \xi)$ is the quadratic Casimir operator for the six-dimensional rotational group whose eigenfucntions are hyperspherical harmonics satisfying 
\begin{equation}
L^{2}(\Omega_{\rho}, \Omega_{\lambda}, \xi) Y_{[\gamma]l_{\rho}l_{\lambda}}(\Omega_{\rho}, \Omega_{\lambda}, \xi) = -\gamma(\gamma + 4 )Y_{[\gamma]l_{\rho}l_{\lambda}}(\Omega_{\rho}, \Omega_{\lambda}, \xi)
\end{equation}
$\gamma = 2n + l_{\rho} + l_{\lambda}$ is the grand angular quantum number. Thus the  hyper-radial part of the wave-function as determined by hypercentral Schrodinger equation is
\begin{equation}
\left[\frac{d^{2}}{dx^{2}} + \frac{5}{x}\frac{d}{dx} - \frac{\gamma(\gamma +4)}{x^{2}}\right]\psi(x) = -2m[E-V(x)]\psi(x)
\end{equation}
Here $l(l+1) \rightarrow \frac{15}{4} + \gamma (\gamma + 4)$.
The hyperradial wave-function $\psi(x)$ is completely symmetric for exchange of  the quark coordinates using the orthogonal basis \cite{giannini-rev}. 
The expansion of quark interaction term as
\begin{equation}
\sum_{i<j}V(r_{ij})= V (x) + ...
\end{equation}

The potential with first term itself gives the hypercentral approximation, which has three-body character as not a single pair of coordinates can be disentangled from the third one. 
The Hamiltonian of the system is written with potential term solely depending on hyperradius {\bf {\it x}} of three body systems. 
\begin{equation}
H = \frac{P^{2}}{2m} + V^{0}(x)  + V_{SD}(x)
\end{equation}
where $m=\frac{2m_{\rho}m_{\lambda}}{m_{\rho}+m_{\lambda}}$ is the reduced mass.
 The potential is solely hyperradius dependent. So, it consists of a Coulomb-like term and a linear term acting as confining part.
\begin{equation}
 V^{0}(x) = -\frac{\tau}{x} + \alpha x
\end{equation}
Here, $\tau = \frac{2}{3}\alpha_{s}$ with $\alpha_{s}$ being running coupling constant.

\begin{equation}
\alpha_{s}= {\frac{\alpha_{s}(\mu_{0})}{1+ {(\frac{33-2 n_{f}}{12\pi})} \alpha_{s}(\mu_{0}) ln{(\frac{m_{1}+m_{2}+m_{3}}{\mu_{0}})}}}
\end{equation}

Here, $\alpha_{s}$ is 0.6 at $\mu_{0}=1 GeV$ and $n_{f}$ is the number of active quark flavors whose value here is 3. $\alpha$ is string tension of the confinement part of the potential. Also, $\alpha$ is state dependent and is obtained by fixing the value using experimental ground state mass of the baryon \cite{zalak16,zalakcpc}. The constituent quark mass is taken to be $m_{s}$ = 0.500 GeV.

\begin{table*}
\centering
\caption{Ground state model parameters}
\begin{tabular}{cccccc}
\hline
$m_{s}$(GeV) & $\alpha_{s}$ &  $\alpha(GeV^{2})$ \\
\hline
0.500 & 0.5109 & 0.0129 \\
\hline
\end{tabular}
\end{table*}
If considering the chiral quark model, the low-energy regime shall be well established as the spontaneously broken SU(3) chiral symmetry scale is different from that of confinement scale of QCD. For three body higher excited states, the relative position of positive and negative parity states can be fixed by the interplay of relativistic kinematics and pion exchange interaction, playing the role of one-gluon exchange potential. Thus, the higher terms in Goldstone exchange will allow us to incorporate the hyperfine, and spin-singlet and triplet splitting \cite{yang, fernandez, glozman}.\\

The $V_{SD}(x)$ is added for incorporating spin-dependent contributions through $V_{SS}(x)$, $V_{\gamma S}(x)$ and $V_{T}(x)$ as spin-spin, spin-orbit and tensor terms respectively. These interactions arise due to $\frac{\nu^{2}}{c^{2}}$ effects in non-relativistic expansion and by the standard Breit-Fermi expansion as described by Voloshin \cite{voloshin}. 

\begin{equation}
\begin{split}
V_{SD}(x) = V_{SS}(x)({\bf S_{\rho}\cdot S_{\lambda}}) +  V_{\gamma S}(x)({\bf \gamma \cdot S}) \\
+ V_{T} [S^{2}- \frac{3({\bf S\cdot x})({\bf S\cdot x})}{x^{2}}]
\end{split} 
\end{equation}

\begin{equation}
V_{SS}(x)=\frac{1}{3m_{\rho}m_{\lambda}}\nabla^{2}V_{V} 
\end{equation}

\begin{equation}
V_{\gamma S}(x)=\frac{1}{2m_{\rho}m_{\lambda}x}(3\frac{dV_{V}}{dx}-\frac{dV_{S}}{dx})
\end{equation}

\begin{equation}
V_{T}(x)=\frac{1}{6m_{\rho}m_{\lambda}}(3\frac{d^{2}V_{V}}{dx^{2}}-\frac{1}{x}\frac{dV_{V}}{dx})
\end{equation}
where $V_{V} = \frac{\tau}{x}$ and $V_{S} = \alpha x$ are the vector and scalar part of potential.
However, in place of spin-spin interaction presented by delta function, we have employed a smear function of the form, details of which can be found in previous works \cite{garcilazo, zalak, kaushal}. Also,  ${\bm  S= \bm S_{\rho} + \bm S_{\lambda}}$ where $\bm S_{\rho}$ and $\bm S_{\lambda}$ are the spin vector associated with the $\bm \rho$ and $\bm \lambda$ variables.
\begin{equation}
V_{SS}(x)=\frac{-A}{6m_{\rho}m_{\lambda}} \frac{e^{\frac{-x}{x_{0}}}}{xx_{0}^{2}}
\end{equation}
Here, $x_{0}$ being the hyperfine parameter, with value $x_{0}=1$ and A is a state dependent parameter consisting of an arbitrary constant. The form of A is chosen as $A=A_{0}/\sqrt{n+l+\frac{1}{2}}$, wherein the value of $A_{0}=28$ for determining the ground state value ($1S\frac{3}{2}^{+}$ 1672 MeV) as well as other radially excited states. Similarly, the other parameters are determined for obtaining the experimentally known ground state mass. i.e. 1672 MeV in the case of $\Omega$. In addition to this, masses with first order correction as $\frac{1}{m}V^{1}(x)$ are taken into account through
\begin{equation}
V^{1}(x)= -C_{F}C_{A}\frac{\alpha_{s}^{2}}{4x^{2}}
\end{equation}
where $C_{F}=\frac{2}{3}$ and $C_{A}=3$ are Casimir elements of fundamental and adjoint representation. \\
Numerical solutions of Schrodinger equation has been obtained through Mathematica notebook \cite{lucha}.

\section{Results and Discussion for the Resonance Mass Spectra}
In the present work, 1S-4S, 1P-4P, 1D-3D and 1F-2F states have been obtained for $S=\frac{1}{2}$ and $S=\frac{3}{2}$ spin configurations with all possible $J^{P}$ values in tables \ref{tab:s-wave} to \ref{tab:f-state}. Also, $Mass_{cal}1$ and $Mass_{cal}2$ signify the resonance masses for without and with first order correction term. Tables \ref{tab:positive} and \ref{tab:negative} give a comparison of obtained results with various approaches for positive and negative parity states. The ground state $\Omega$(1672) is nearly same for many the approaches with a variation of 20-30 MeV in few cases depending on the approach. \\

Faustov {\it et al.} \cite{faustov} has employed a relativistic quark model approach considering quark-diquark system. The lower excited states are very much in accordance but for higher excitations exact comparison is not possible. In ref \cite{ma} has studied the spectrum through hyperfine interactions due to two-gluon exchange. For the available states, the present results are very near within 50 MeV compared to theirs. Another non-relativistic constituent quark model approach has been utilized by \cite{liu}. Y. Oh \cite{oh} has investigated with $\Omega$ spectrum using Skyrme model. \cite{capstick} and \cite{chao} have exploited quark model based on chromodynamics and some of the present states are also in accordance. E. Klempt \cite{klempt} has reproduced the few known states through a new baryon mass formula. U. L\"{o}ring {\it et al.} has studied the whole light spectrum within relativistic covariant quark model based on Bethe-Salpeter equation \cite{loring}. The BGR collaboration \cite{bgr} results are based on chirally improved (CI) quarks. For higher $J^{P}$ values, not many approaches are available for comparison. \\
One puzzling question remains with $\Omega$(2012) state, however our results donot exactly reproduce but vary by 30 MeV. Also, this study is not able to precisely comment on the proposed molecular nature of this state. So, the future experimental results would serve as a key towards its understanding. \\

The results described in tables \ref{tab:s-wave} to \ref{tab:f-state} have been summarized in the increasing order for each $J^{P}$ value including positive parity in table \ref{tab:positive} and negative parity in table \ref{tab:negative}. All the mentioned models appearing in the table for comparison are not sufficient to segregate each state based on $J^{P}$ value. For the ground state $\Omega$(1672) with $J^{P}=\frac{3}{2}^{+}$ is near to \cite{faustov}, \cite{ma}, \cite{liu} and \cite{chao}. The first state with $J^{P}=\frac{1}{2}^{+}$ is nearly comparable to the relativistic approach by Faustov et. al. 

\end{multicols}

\begin{table}
\begin{minipage}{0.45\linewidth}
\vspace{-2.5cm}
\centering
\caption{ Resonance masses of S-state 1S-4S for without and with first order correction to the potential (in MeV)}
\label{tab:s-wave}
\begin{tabular*}{\linewidth}{@{\extracolsep{\fill}}cccc}
\hline
State & $J^{P}$ &  $Mass_{cal}$1 & $Mass_{cal}$2  \\
\hline
1S & $\frac{3}{2}^{+}$ & 1672 & 1672  \\
2S & $\frac{3}{2}^{+}$ & 2057 & 2068  \\
3S & $\frac{3}{2}^{+}$ & 2429 & 2449  \\
4S & $\frac{3}{2}^{+}$ & 2852 & 2885  \\
\hline
\end{tabular*}

\vspace{0.5cm}
\caption{ Resonance masses of P-state 1P-4P for without and with first order correction to the potential (in MeV)}
\label{tab:p-wave}
\begin{tabular*}{\linewidth}{@{\extracolsep{\fill}}cccc}
\hline
State & $J^{P}$ &  $Mass_{cal}$1 & $Mass_{cal}$2  \\
\hline
$1^{2}P_{1/2}$ & $\frac{1}{2}^{-}$ & 1987 & 1996 \\
$1^{2}P_{3/2}$ & $\frac{3}{2}^{-}$ & 1978 & 1985 \\
$1^{4}P_{1/2}$ & $\frac{1}{2}^{-}$ & 1992 & 2001 \\
$1^{4}P_{3/2}$ & $\frac{3}{2}^{-}$ & 1983 & 1991 \\
$1^{4}P_{5/2}$ & $\frac{5}{2}^{-}$ & 1970 & 1997 \\
\hline
$2^{2}P_{1/2}$ & $\frac{1}{2}^{-}$ & 2345 & 2363 \\
$2^{2}P_{3/2}$ & $\frac{3}{2}^{-}$ & 2332 & 2349 \\
$2^{4}P_{1/2}$ & $\frac{1}{2}^{-}$ & 2352 & 2370 \\
$2^{4}P_{3/2}$ & $\frac{3}{2}^{-}$ & 2339 & 2356 \\
$2^{4}P_{5/2}$ & $\frac{5}{2}^{-}$ & 2321 & 2338 \\
\hline
$3^{2}P_{1/2}$ & $\frac{1}{2}^{-}$ & 2758 & 2788 \\
$3^{2}P_{3/2}$ & $\frac{3}{2}^{-}$ & 2740 & 2770 \\
$3^{4}P_{1/2}$ & $\frac{1}{2}^{-}$ & 2767 & 2797 \\
$3^{4}P_{3/2}$ & $\frac{3}{2}^{-}$ & 2749 & 2779 \\
$3^{4}P_{5/2}$ & $\frac{5}{2}^{-}$ & 2726 & 2755 \\
\hline
$4^{2}P_{1/2}$ & $\frac{1}{2}^{-}$ & 3218 & 3264 \\
$4^{2}P_{3/2}$ & $\frac{3}{2}^{-}$ & 3196 & 3240 \\
$4^{4}P_{1/2}$ & $\frac{1}{2}^{-}$ & 3229 & 3276 \\
$4^{4}P_{3/2}$ & $\frac{3}{2}^{-}$ & 3207 & 3252 \\
$4^{4}P_{5/2}$ & $\frac{5}{2}^{-}$ & 3178 & 3221 \\
\hline
\end{tabular*}
\end{minipage}%
\hspace{1cm}
\begin{minipage}{0.45\linewidth}
\centering
\caption{ Resonance masses of D-state 1D-3D for without and with first order correction to the potential (in MeV)}
\label{tab:d-state}
\begin{tabular*}{\linewidth}{@{\extracolsep{\fill}}cccc}
\hline
State & $J^{P}$ & $Mass_{cal}$1 & $Mass_{cal}$2 \\
\hline
$1^{2}D_{3/2}$ & $\frac{3}{2}^{+}$ & 2269 & 2288 \\
$1^{2}D_{5/2}$ & $\frac{5}{2}^{+}$ & 2250 & 2267 \\
$1^{4}D_{1/2}$ & $\frac{1}{2}^{+}$ & 2291 & 2311 \\
$1^{4}D_{3/2}$ & $\frac{3}{2}^{+}$ & 2276 & 2295 \\
$1^{4}D_{5/2}$ & $\frac{5}{2}^{+}$ & 2257 & 2275 \\
$1^{4}D_{7/2}$ & $\frac{7}{2}^{+}$ & 2233 & 2249 \\
\hline
$2^{2}D_{3/2}$ & $\frac{3}{2}^{+}$ & 2671 & 2703 \\
$2^{2}D_{5/2}$ & $\frac{5}{2}^{+}$ & 2646 & 2676 \\
$2^{4}D_{1/2}$ & $\frac{1}{2}^{+}$ & 2699 & 2733 \\
$2^{4}D_{3/2}$ & $\frac{3}{2}^{+}$ & 2681 & 2713 \\
$2^{4}D_{5/2}$ & $\frac{5}{2}^{+}$ & 2656 & 2686 \\
$2^{4}D_{7/2}$ & $\frac{7}{2}^{+}$ & 2623 & 2652 \\
\hline
$3^{2}D_{3/2}$ & $\frac{3}{2}^{+}$ & 3122 & 3166 \\
$3^{2}D_{5/2}$ & $\frac{5}{2}^{+}$ & 3092 & 3135 \\
$3^{4}D_{1/2}$ & $\frac{1}{2}^{+}$ & 3157 & 3201 \\
$3^{4}D_{3/2}$ & $\frac{3}{2}^{+}$ & 3134 & 3178 \\
$3^{4}D_{5/2}$ & $\frac{5}{2}^{+}$ & 3103 & 3146 \\
$3^{4}D_{7/2}$ & $\frac{7}{2}^{+}$ & 3065 & 3107 \\
\hline
\end{tabular*}
\vspace{0.5cm}
\caption{ Resonance masses of F-state 1F-2F for without and with first order correction to the potential (in MeV)}
\label{tab:f-state}
\begin{tabular*}{\linewidth}{@{\extracolsep{\fill}}cccc}
\hline
State & $J^{P}$ &  $Mass_{cal}$1 & $Mass_{cal}$2  \\
\hline
$1^{2}F_{5/2}$ & $\frac{5}{2}^{-}$ & 2585 & 2614 \\
$1^{2}F_{7/2}$ & $\frac{7}{2}^{-}$ & 2552 & 2579 \\
$1^{4}F_{3/2}$ & $\frac{3}{2}^{-}$ & 2622 & 2653 \\
$1^{4}F_{5/2}$ & $\frac{5}{2}^{-}$ & 2595 & 2625 \\
$1^{4}F_{7/2}$ & $\frac{7}{2}^{-}$ & 2562 & 2590 \\
$1^{4}F_{9/2}$ & $\frac{9}{2}^{-}$ & 2521 & 2548 \\
\hline
$2^{2}F_{5/2}$ & $\frac{5}{2}^{-}$ & 3027 & 3069 \\
$2^{2}F_{7/2}$ & $\frac{7}{2}^{-}$ & 2986 & 3027 \\
$2^{4}F_{3/2}$ & $\frac{3}{2}^{-}$ & 3072 & 3115 \\
$2^{4}F_{5/2}$ & $\frac{5}{2}^{-}$ & 3039 & 3081 \\
$2^{4}F_{7/2}$ & $\frac{7}{2}^{-}$ & 2998 & 3040 \\
$2^{4}F_{9/2}$ & $\frac{9}{2}^{-}$ & 2949 & 2999 \\
\hline
\end{tabular*}
\end{minipage}
\end{table}

\begin{table*}
\centering
\caption{Comparison of present masses with other approaches based on $J^{P}$ value with positive parity described in the increasing order for all possible spin-parity assignment (in MeV)}
\label{tab:positive}
\begin{tabular}{ccccccccccccc}
\hline
$J^{P}$ & $Mass_{cal}1$ & $Mass_{cal}2$ & \cite{faustov} & \cite{ma} & \cite{liu} & \cite{pervin} & \cite{oh} & \cite{capstick} & \cite{chao} & \cite{loring} & \cite{bgr}\\
\hline
$\frac{1}{2}^{+}$ & 2291 & 2311 & 2301 & 2182 & 2232 & 2175 & 2140 & 2220 & 2190 & 2232 & 2350(63)\\ 
  & 2699 & 2733 & & 2202 & & 2191 & & 2255 & 2210 & 2256 &  2481(51)\\
  & 3157 & 3201 & & & & & & & & & \\
\hline
$\frac{3}{2}^{+}$ & 1672 & 1672 & 1678 & 1673 & 1672 & 1656 & 1694 & 1635 & 1675 & & 1642(17)\\
 & 2057 & 2068 & 2173 & 2078 & 2159 & 2170 & 2282 & 2165 & 2065 & 2177 & 2470(49)\\
 & 2269 & 2288 & 2304 & 2208 & 2188 & 2182 & & 2280 & 2215 & 2236 & &\\
 & 2276 & 2295 & 2332 & 2263 & 2245 & & & 2345 & 2265 & 2287 & &\\
 & 2429 & 2449 & & & & & & & & & &\\
 & 2671 & 2703 & & & & & & & & & &\\
 & 2681 & 2713 & & & & & & & & & &\\
 & 2852 & 2885 & & & & & & & & & &\\
 & 3122 & 3166 & & & & & & & & & &\\
 & 3134 & 3178 & & & & & & & & & &\\
\hline
$\frac{5}{2}^{+}$ & 2250 & 2267 & 2401 & 2224 & 2303 & 2178 & & 2280 & 2225 & 2253 & & \\
 & 2257 & 2275 & & 2260 & 2252 & 2210 & & 2345 & 2265 & 2312 & & \\
 & 2646 & 2676 & & & & & & & & & & \\
 & 2656 & 2686 & & & & & & & & & & \\
 & 3092 & 3135 & & & & & & & & & & \\
 & 3102 & 3146 & & & & & & & & & & \\
\hline
$\frac{7}{2}^{+}$ & 2233 & 2249 & 2369 & 2205 & 2321 & 2183 & & 2295 & 2210 & 2292 & & \\
 & 2623 & 2652 & & & & & & & & & & \\
 & 3065 & 3107 & & & & & & & & & & \\
 \hline
\end{tabular}
\end{table*}

\begin{table*}
\centering
\caption{Comparison of present masses with other approaches based on $J^{P}$ value with negative parity described in the increasing order for all possible spin-parity assignment (in MeV)}
\label{tab:negative}
\begin{tabular}{cccccccccccccc}
\hline
$J^{P}$ & $Mass_{cal}1$ & $Mass_{cal}2$ & \cite{faustov} & \cite{ma} & \cite{liu} & \cite{pervin} & \cite{oh} & \cite{capstick} & \cite{chao} & \cite{loring} & \cite{bgr} \\
\hline
$\frac{1}{2}^{-}$ & 1987 & 1996 & 1941 & 2015 & 1957 & 1923 & 1837 & 1950 & 2020 & 1992 & 1944(56) \\
 & 1983 & 2001 & 2463 & & & & & 2410 & & 2456 & 2716(118)\\
 & 2345 & 2363 & 2580 & & & & & 2490 & & 2498 & & \\
 & 2352 & 2370 & & & & & & & 2550 & & \\
 & 2758 & 2788 & & & & & & & & & \\
 & 2767 & 2797 & & & & & & & & & \\
 & 3218 & 3264 & & & & & & & & & \\
 & 3229 & 3276 & & & & & & & & & \\
\hline 
$\frac{3}{2}^{-}$ & 1978 & 1985 & 2038 & 2015 & 2012 & 1953 & 1978 & 2000 & 2020 & 1976 & 2049(32)\\
 & 1983 & 1991 & 2537 & & & & 2604 & 2440 & & 2446 & 2755(67)\\
 & 2332 & 2349 & 2636 & & & & & 2495 & & 2507 & & \\
 & 2339 & 2356 & & & & & & & & 2524 & & \\
 & 2622 & 2653 & & & & & & & & 2564 & & \\
 & 2740 & 2770 & & & & & & & & 2594 & & \\
 & 2749 & 2779 & & & & & & & & & & \\
 & 3072 & 3115 & & & & & & & & & & \\
 & 3196 & 3240 & & & & & & & & & & \\
 & 3207 & 3252 & & & & & & & & & & \\
\hline
$\frac{5}{2}^{-}$ & 1970 & 1997 & 2653 & & & & & 2490 & & 2528 & & \\
 & 2321 & 2338 & & & & & & & & 2534 & & \\
 & 2585 & 2614 & & & & & & & & 2554 & & \\
 & 2595 & 2625 & & & & & & & & 2617 & & \\
 & 2726 & 2755 & & & & & & & & & & \\
 & 3027 & 3069 & & & & & & & & & & \\
 & 3039 & 3081 & & & & & & & & & & \\
 & 3178 & 3221 & & & & & & & & & & \\
\hline
$\frac{7}{2}^{-}$ & 2562 & 2590 & 2599 & & & & & & & 2531 & & \\
 & 2998 & 3040 & & & & & & & & 2577 & & \\
\hline
$\frac{9}{2}^{-}$ & 2521 & 2548 & 2649 & & & & & & & 2606 & & \\
 & 2949 & 2999 & & & & & & & & & & \\
\hline
\end{tabular}
\end{table*}

\begin{multicols}{2}

\begin{table*}
\centering
\caption{Regge slopes and intercepts for (n,$M^{2}$)}
\label{tab: slope}
\begin{tabular}{ccc}
\hline
Trajectory & $b$ &  $b_{0}$ \\
 \hline
S  & 1.76838 $\pm$ 0.12837 & 0.84426 $\pm$ 0.35156 \\
P  & 2.07004 $\pm$ 0.1839 & 1.52458 $\pm$ 0.50364 \\
D & 2.20397 $\pm$ 0.17905 & 2.67895 $\pm$ 0.3868 \\
F & 2.34116 & 4.01428 \\
 \hline
\end{tabular}
\end{table*}

\begin{table*}
\centering
\caption{Comparison of ground state magnetic moment (in $\mu_{N}$)}
\label{tab:mm}
\begin{tabular}{cccccccccc}
\hline
Present & Exp & \cite{dhir} & \cite{dhir} & \cite{hong} &\cite{linde} & \cite{sahu} & \cite{aarti} & \cite{harleen} & \cite{fayyazuddin}\\
\hline
-1.68 & -2.02 & -1.67 & -1.90 & -2.06 & -1.95 & -1.61 & -2.08 & -2.01 & -1.84\\
\hline
\end{tabular}
\end{table*}

\newpage
The first excited states with $J^{P}=\frac{5}{2}^{+}$ and $J^{P}=\frac{7}{2}^{+}$ are not very far from most of the comparison. As in the case on negative parity states $J^{P}=\frac{3}{2}^{-}$, \cite{ma}, \cite{liu}, \cite{capstick} and \cite{chao}, the results are near 2012 MeV, whereas present study couldn't identify exactly the proposed $\Omega$(2012) state. The state $J^{P}=\frac{7}{2}^{-}$ is very much near to the results by \cite{faustov} and \cite{loring}. \\

 Here, we attempt to assign a tentative spin-parity to the three and two starred states. The $\Omega$(2250) with experimental mass at $2252 \pm 9$ is quite comparable to our 1D with $J^{P}=\frac{5}{2}^{+}$. The fourth state $\Omega$(2380) may possibly be a member of 2P family with $J^{P}=\frac{1}{2}^{-}$ matching our value at 2370 MeV. The last state of $\Omega$(2470) with mass as $2474 \pm 12$ might be assigned a 3S state with $J^{P}=\frac{3}{2}^{+}$ as the calculated state 2429 or 2449 MeV. \\

The present model has attempted to distinguish all the possible spin-parity assignment of the excited states however, due to limited data obtained by various compared models, exact state-wise comparison is not possible. Thus, this study is expected to provide a possible range of masses for upcoming experiments which shall identify the existence of a particular state.

\section{Regge Trajectories}
\begin{figure*}
\centering
\caption{\label{fig:omega-nm} {Regge trajectory $n \rightarrow M^{2}$ for S, P, D and F state masses and linearly fitted. }}
\includegraphics[scale=0.4]{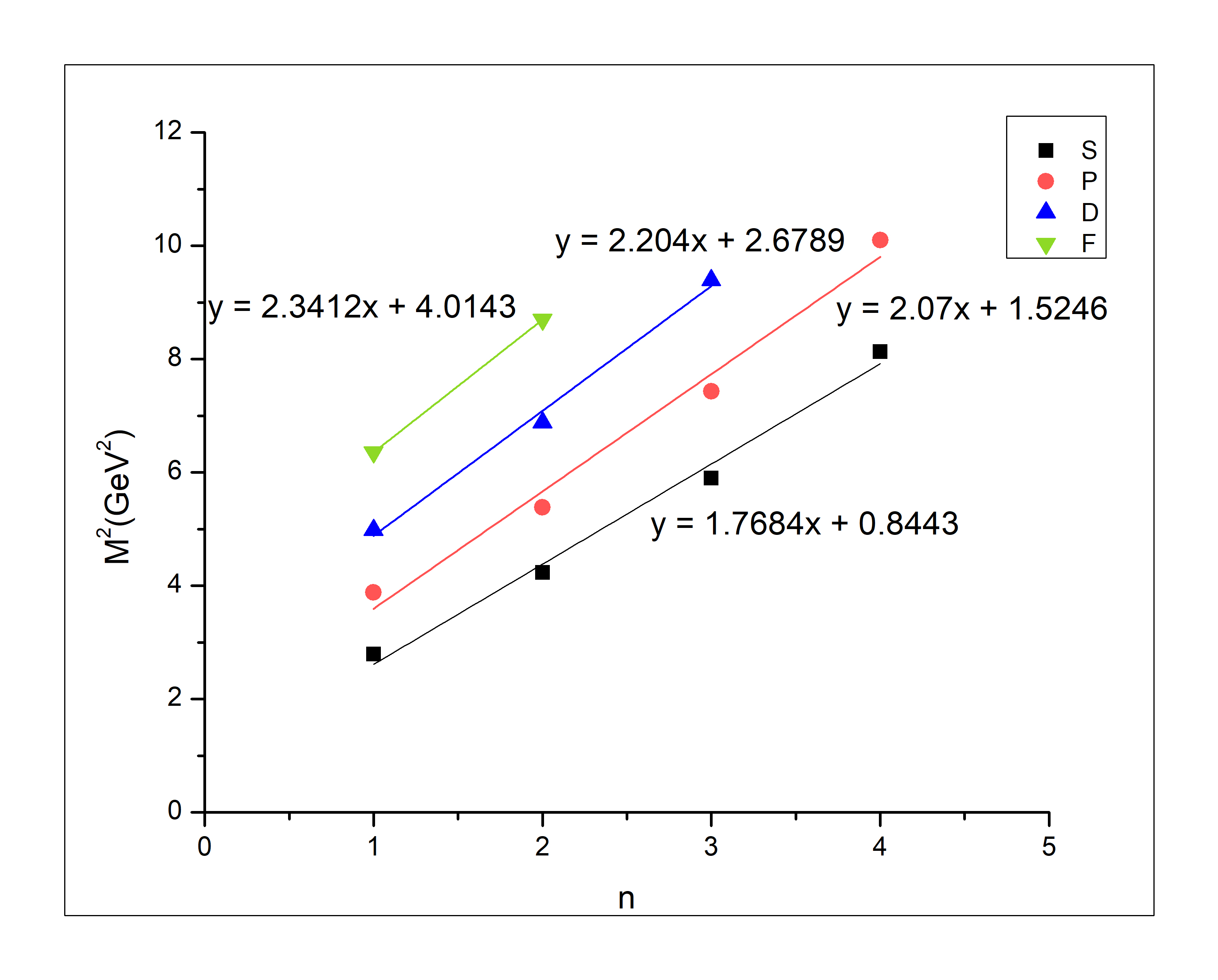}
\end{figure*}

\begin{figure*}
\centering
\caption{\label{fig:omega-jm12}{Regge trajectory $J \rightarrow M^{2}$ for natural parity} }
\includegraphics[scale=0.4]{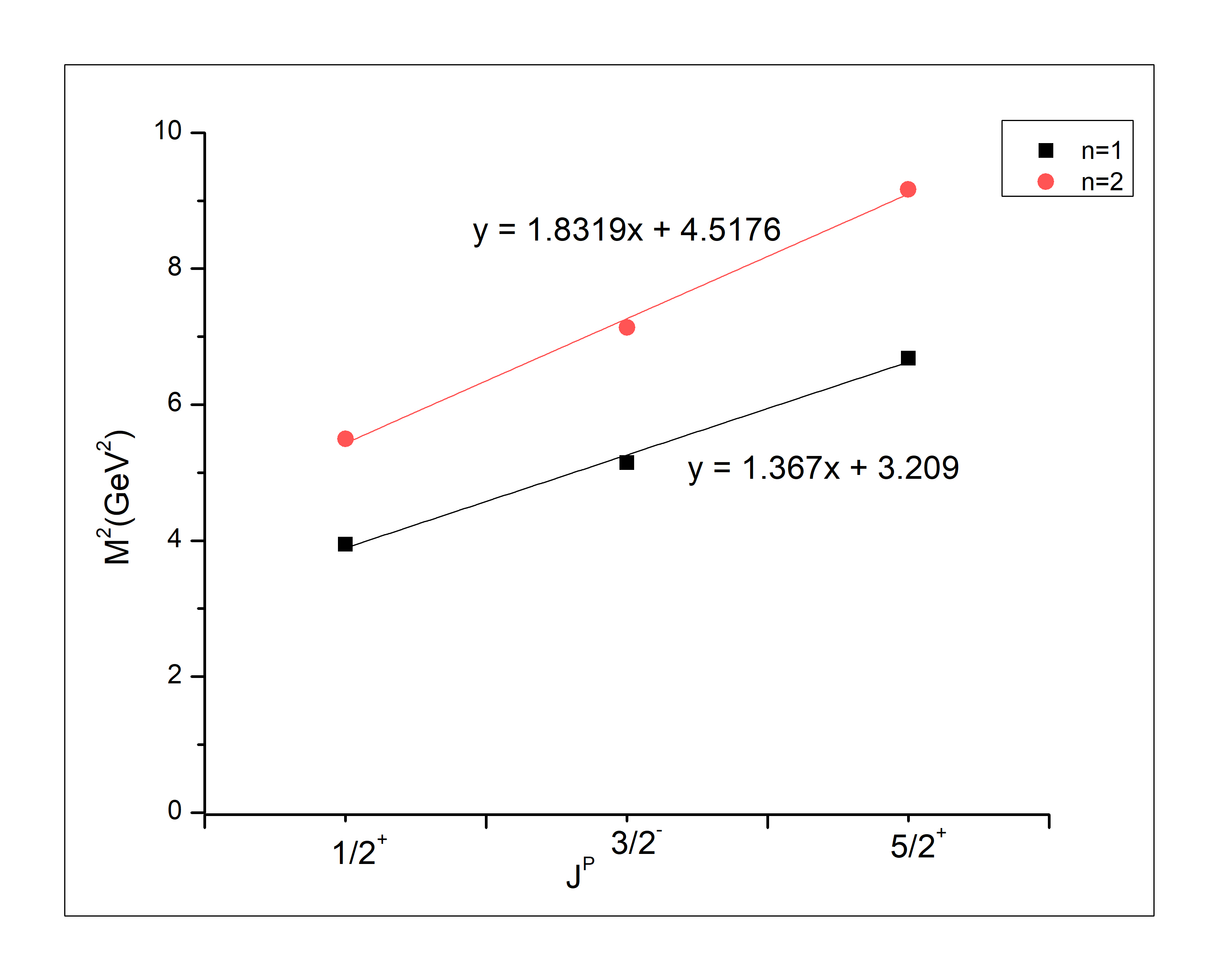}
\end{figure*}

\begin{figure*}
\centering
\caption{\label{fig:omega-jm32} {Regge trajectory $J \rightarrow M^{2}$ for unnatural parity }}
\includegraphics[scale=0.4]{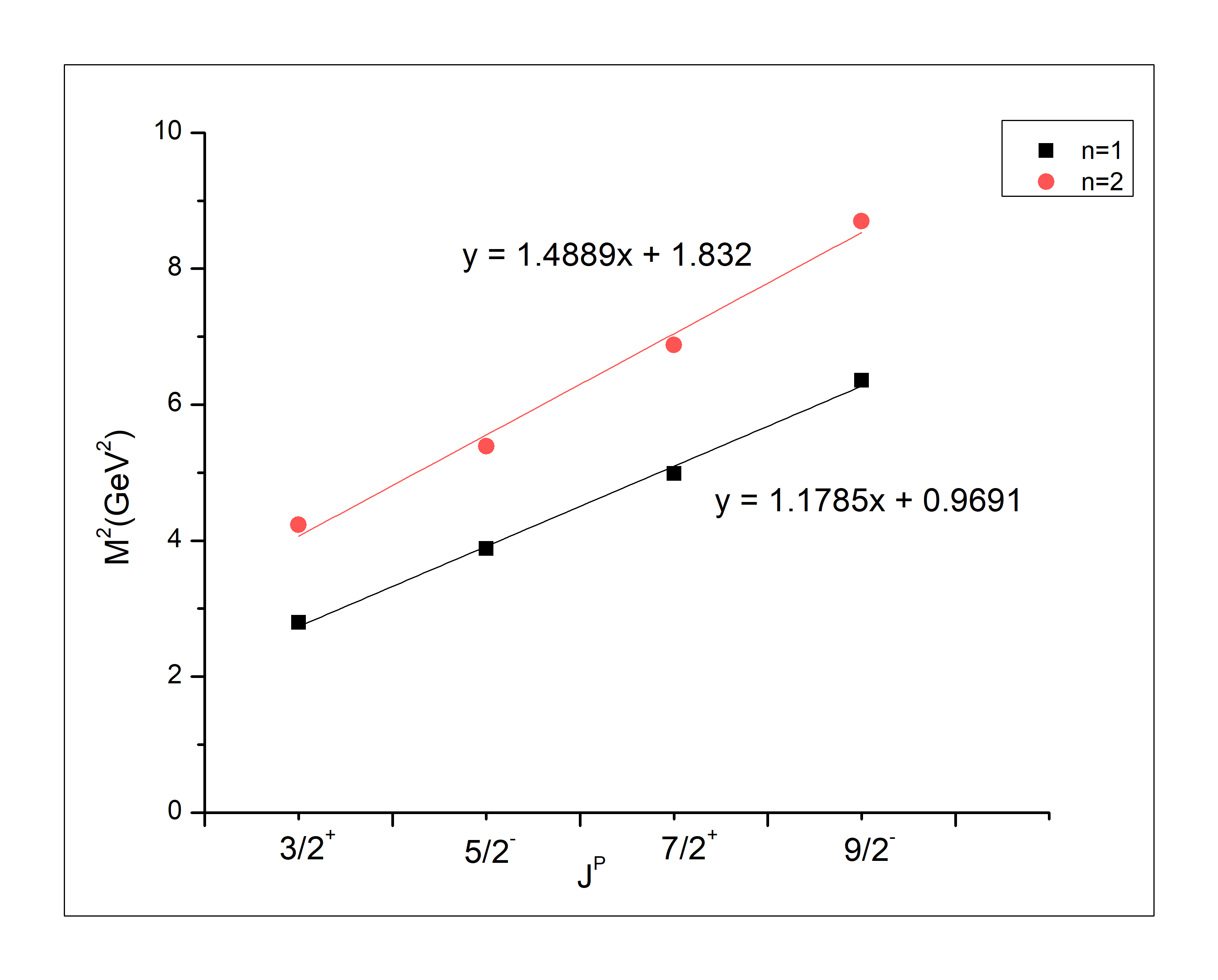}
\end{figure*}

Regge trajectories have been of importance in spectroscopic studies. The plot of total angular momentum J and principal quantum number n against the square of resonance mass $M^{2}$ are drawn to obtain the non-intersecting and linearly fitted lines. Figure \ref{fig:omega-nm} shows a linear behaviour with almost all the points following the trend for $n-M^{2}$. Figures \ref{fig:omega-jm12} and \ref{fig:omega-jm32} are plotted with few natural and unnatural parity states for available results. These plots point that the spin-parity assignment of a given state in present calculation could possibly be correct. The linear fitting parameters are mentioned in the respective plots. 
\begin{subequations}
\begin{align}
J = aM^{2} + a_{0} \\
n = b M^{2} + b_{0}
\end{align}
\end{subequations}

Faustov {\it et al.} \cite{faustov} has shown the total angular momentum J against the square of mass trajectories using the resonances obtained with relativistic quark model. The slope and intercept values has been given as 0.712 $\pm$ 0.002 and -0.504 $\pm$ 0.007 respectively. The similar plot with natural parity for present masses gives the value as 1.1785 $\pm$ 0.048 and 0.9691 $\pm$ 0.154 respectively. The values for slope and intercept with standard error for (n,$M^{2}$) plot are listed in table \ref{tab: slope}. However, due to lack of more experimental data, we are unable to comment on exact comparison of the values. 

%
%

%
%

%

\section{Magnetic Moment}
The electromagnetic properties of baryons are a challenging realm especially for short-lived $J^{P}=\frac{3}{2}^{+}$ decuplet baryons. Many theoretical approaches have attempted to investigate strange baryon magnetic moments. It serves as an asset for the study of decay properties as well as intrinsic dynamics of quarks. The generalized form of magnetic moment is  \cite{kaushal}
\begin{equation}
\mu_{B}= \sum_{q} \left\langle \phi_{sf} \vert \mu_{qz}\vert \phi_{sf} \right\rangle
\end{equation}
where $\phi_{sf}$ is the spin-flavour wave function. The contribution from individual quark appears as 
\begin{equation}
\mu_{qz}= \frac{e_{q}}{2m_{q}^{eff}}\sigma_{qz}
\end{equation}
$e_{q}$ being the quark charge, $\sigma_{qz}$ being the spin orientation and $m_{q}^{eff}$ is the effective mass which may vary from model based quark mass due to interactions. In case of $\Omega$, $\sigma_{qz}=s \uparrow s \uparrow s \uparrow$ which leads to $3\mu_{s}$. Table \ref{tab:mm} summarizes the calculated magnetic moment alongwith other comparison results \cite{dhir,hong,linde,sahu,aarti,harleen,fayyazuddin}.

\section{Conclusion}
A hypercentral Constituent Quark Model with a linear confining term, spin-dependent terms and correction term has been helpful to obtain mass spectra upto higher excited states upto nearly 3 GeV. Even though the scarcity of experimental data doesnot allow us to completely validate the findings but comparison with theoretical models with varied assumptions are of keen interest. The low-lying states are quite in accordance with some models but not exactly matching for higher $J^{P}$ values.\\
 
It is noteworthy that the current findings could not comment on the debated state of $\Omega$(2012) for molecular structure, however the mass varies within 30 MeV with $J^{P}=\frac{3}{2}^{-}$ which may be identified as a negative parity state of 1P family.  As the $J^{P}$ value for any other state is not experimentally known, exact comparison still depends upon more future findings. However, the probable spin-parity assignment according to the obtained value can be given. Thus, $\Omega$(2250) could be $1D_{\frac{5}{2}^{+}}$ state; $\Omega$(2380) be $2P_{\frac{1}{2}^{-}}$. and $\Omega$(2470) may be associated with 3S with $\frac{3}{2}^{+}$.  \\

The magnetic moment differs by 0.5$\mu_{n}$ from PDG and other results. The Regge trajectories show the linear nature giving the hint for spin-parity assignments could possibly be correct. However, the validation of any of the results depends on the future experimental facilities to exclusively study the strange baryon properties especially by $\bar P$ANDA at FAIR-GSI \cite{panda} and BESIII \cite{BESIII}.

\section*{Acknowledgement}
Ms. Chandni Menapara would like to acknowledge the support from the Department of Science and Technology (DST) under INSPIRE-FELLOWSHIP scheme for pursuing this work.\\
  

\end{multicols}

\clearpage
\end{CJK*}
\end{document}